\documentclass[twocolumn, superscriptaddress,showpacs,amsmath,amssymb]{revtex4}
\usepackage{mathrsfs}
\usepackage{hyperref}

\usepackage[dvips]{graphics,epsfig}
\usepackage{amstext}
\usepackage{amssymb}
\usepackage{amsmath}
\usepackage{graphicx,color}
\usepackage{bm}

\newcommand{\s}{\mbox{\tiny S}}
\newcommand{\tS}{\mbox{\tiny S}}

\newcommand{\B}{\mbox{\tiny B}}
\newcommand{\U}{\mbox{\tiny U}}
\newcommand{\SB}{\mbox{\tiny SB}}

\newcommand{\M}{\mbox{\tiny M}}

\newcommand{\tL}{\mbox{\tiny L}}
\newcommand{\tR}{\mbox{\tiny R}}
\newcommand{\ti}{\Tilde}
\newcommand{\wti}{\widetilde}
\newcommand{\nl}{\nonumber \\}

\newcommand{\Sec}[1]{Sec.\,\ref{#1}}

\newcommand{\be}{\begin{equation}}
\newcommand{\ee}{\end{equation}}
\newcommand{\bea}{\begin{eqnarray}}
\newcommand{\eea}{\end{eqnarray}}
\newcommand{\bsube}{\begin{subequations}}
\newcommand{\esube}{\end{subequations}}
\newcommand{\ben}{\begin{equation*}}
\newcommand{\een}{\end{equation*}}
\newcommand{\Eq}[1]{Eq.\,(\ref{#1})}

\newcommand{\Fig}[1]{Fig.\,\ref{#1}}

\newcommand{\dg}{\dagger}
\newcommand{\la}{\langle}
\newcommand{\ra}{\rangle}

\newcommand{\La}{\big\la}
\newcommand{\Ra}{\big\ra}

\newcommand{\Opm}{\hat O_{\mbox{\tiny $\pm$}}}
\newcommand{\half}{\mbox{ $\frac{1}{2}$}}

\begin{document}
 
 
\title{Coulomb interaction unlocks Majorana-mediated electron teleportation between Quantum dots}

\author{Sirui Yu}
\affiliation{School of Physics, Hangzhou Normal University,
Hangzhou, Zhejiang 311121, China}

\author{Hong Mao}  
\affiliation{School of Physics, Hangzhou Normal University,
 Hangzhou, Zhejiang 311121, China}
 
 \author{Jinshuang Jin} \email{jsjin@hznu.edu.cn}
\affiliation{School of Physics, Hangzhou Normal University,
Hangzhou, Zhejiang 311121, China}

 \author{Chui-Ping Yang} 
\affiliation{School of Physics, Hangzhou Normal University,
Hangzhou, Zhejiang 311121, China}

\date{\today}

\begin{abstract}
 We investigate quantum transport in a hybrid system 
 composed of two quantum dots (QDs) 
  coupled through a pair of spatially separated Majorana zero modes (MZMs)
 with negligible coupling energy. 
We focus on nonlocal correlations mediated by the MZMs, 
 particularly the role of Coulomb interaction $U$ 
 between the QDs and the Majorana wire.  
 Using the numerically exact 
 fermionic dissipation equation of motion (DEOM) method, 
 we compute both the transient current and the current-current cross-correlation noise spectrum.
In the non-interacting case ($U=0$), 
destructive interference between the degenerate normal tunneling 
and anomalous tunneling channels suppresses 
electron teleportation between the dots. 
Introducing a finite Coulomb interaction $U$ lifts this channel degeneracy, 
enabling strong nonlocal correlations and 
 inter-dot electron teleportation. 
This effect manifests as a robust signal in the cross-correlation noise spectrum, 
  which is significantly stronger than that induced by 
 a finite Majorana coupling energy $\varepsilon_{\M}$.
  Our findings propose Coulomb interaction as an efficient and experimentally 
  accessible control parameter for generating and detecting Majorana-mediated 
  nonlocal transport in the topologically relevant 
  long-wire limit ($\varepsilon_{\M}\rightarrow0$).

\end{abstract}

\maketitle

\section{Introduction}

The pursuit of topological quantum computation has stimulated intense 
interest in Majorana zero 
modes (MZMs), as evidenced by extensive theoretical 
\cite{Fle10180516,Lut10077001,Ore10177002,Law09237001,Liu13064509,
  Ulr15075443,Bol07237002,Nil08120403,Zoc13036802,  
  Bjo13036802,Lu14195404,Lu16245418, Fen21123032}
and experimental 
\cite{Mou121003,Den126414,Das12887,  
Fin13126406,Alb161038} studies.
MZMs are exotic quasiparticles obeying non-Abelian statistics, 
 predicted to emerge at the ends of one-dimensional 
topological superconducting nanowires \cite{Lut10077001,Ore10177002}.
Their nonlocal nature, manifested as spatially separated zero-energy 
states, offers a promising platform for fault-tolerant quantum 
information processing \cite{Kit01131,Kit032,Nay081083,Lei11210502}. 
A key signature of this nonlocality is 
 electron teleportation 
\cite{Sem071479,Tew08027001,Fu10056402},
whereby an electron is coherently transferred between distant points via MZMs 
without direct overlap of the electronic wavefunctions.


A common approach for demonstrating nonlocality 
involves the measurement of nonlocal current cross-correlations \cite{Law09237001,Liu13064509,
  Ulr15075443,Bol07237002,Nil08120403,Zoc13036802, 
  Bjo13036802,Lu14195404,Lu16245418, Fen21123032}. 
Previous work have shown that a finite Majorana coupling energy  
($\varepsilon_{\M}$) can generate non-zero cross-correlation noise.
However, this noise vanishes in the limit $\varepsilon_{\M}\rightarrow0$,
both for MZMs coupled directly to leads \cite{Bol07237002,Nil08120403,Zoc13036802}
and for MZMs coupled via QDs (referred to as 
a QD-MZM system) \cite{Bjo13036802,Lu14195404,Lu16245418,
Fen21123032}, as illustrated in \Fig{fig1}.
%
%
Notably, Feng {\it et al.,} \cite{Fen21123032} clarified that the absence of
cross-correlation at $\varepsilon_{\M}\rightarrow0$ stems from a degeneracy of 
between the
Andreev process and the `teleportation', channels. 
They proposed lifting this degeneracy  
by introducing finite charging energy $E_C$ on the Majorana island, 
which restores nonzero cross-correlation 
even in the $\varepsilon_{\M}\rightarrow0$ limit within the QD-MZM 
transport geometry \cite{Fen21123032}.
This work generalizes Ref.\,\cite{Fu10056402}, where a large $E_C$ limit 
was suggested to
  fully suppress the Andreev process, thereby isolating a 
  pure electron `teleportation' channel alone 
 for MZMs coupled directly to leads.  
 The role of large $E_C$ in restoring cross-correlation noise
  at $\varepsilon_{\M}\rightarrow0$ has also been demonstrated
  in 
Ref.\,\cite{Lu16245418}.
 Furthermore, nonlocal properties arising from 
 finite $\varepsilon_{\M}$ (via shorter wires) or  
a finte/large charging energy $E_C$ 
have been proposed to generate entanglement between
QDs \cite{Zhi13214513,Ste15214507,Vim24224510,Jas25075415}  
or nanomechanical oscillators \cite{Ste14155431}.
Nevertheless, $\varepsilon_{\M}$ is inherently 
small in long wires, which are desirable for topological protection, 
and experimental detection of nonlocal correlations remains elusive..

In this work, we propose and theoretically demonstrate that 
Coulomb interaction between the QDs and the Majorana wire provides 
an efficient and experimentally viable alternative mechanism to 
realize the electron teleportation
 between two well-separated QDs mediated by MZMs 
in a long topological wire ($\varepsilon_{\M}\rightarrow0$).
 As revealed in Ref.\,\cite{Ric20165104}, Coulomb interaction between the QD
and the superconducting section induces an additional overlap
of the wave functions describing the individual MZMs,
even when they are spatially distant.
For InAs systems, the interaction strength $U$ is 
estimated to range 
from ${\rm \mu eV}$ to ${\rm meV}$, 
depending on the size of the QD
formed at the bare region nanowire \cite{Ric20165104}.

Reanalyzing the intrinsic transport mechanism for
 the hybrid QD-MZM system (\Fig{fig1}), 
we find that it is
described by two distinct quantum 
tunneling processes: normal tunneling (NT) 
and anomalous tunneling (AT). These are associated with 
electron ($\varepsilon_{\M}$) and 
hole ($-\varepsilon_{\M}$) transport channels  
 via the Majorana fermionic mode, respevtively.
In the ideal long-wire limit ( $\varepsilon_{\M}\rightarrow 0$), 
these two tunneling channels become degenerate. This degeneracy 
causes destructive quantum interference between the NT and AT 
paths, suppressing the coherent transfer of 
electrons between the dots and eliminating nonlocal cross-correlations. 
We show that
 Coulomb interaction between QDs and Majorana wire 
breaks this channel degeneracy, thereby unlocking robust Majorana-mediated electron teleportation.
 Using the numerically exact fermionic dissipation equation of motion (DEOM) method, 
 we investigate quantum transport through the QD-MZM structure (\Fig{fig1}). 
 We compute both transient currents $I(t)$ and the 
current–current cross-correlation noise spectrum $S_{\tL\tR}(\omega)$,
presenting their characteristic features in detail.

This paper is organized as follows. In \Sec{thmeth}, we describe the model Hamiltonian and 
outline the DEOM methodology for accurate quantum transport calculations. 
Section \ref{thham} analyzes the isolated hybrid QD–MZM system, 
contrasting the non-interacting 
and interacting cases to elucidate the role of $U$ in establishing
inter-dot correlations. In \Sec{thmea}, we present numerical 
results for transport currents and the cross-correlation noise spectrum, 
demonstrating how Coulomb interaction generates measurable 
teleportation signatures. Finally, we summarize our 
findings in \Sec{thsum}.


\section{Methodology}
\label{thmeth}

\subsection{Model description}
\label{thmod}

\begin{figure}
\includegraphics[width=0.8\columnwidth]{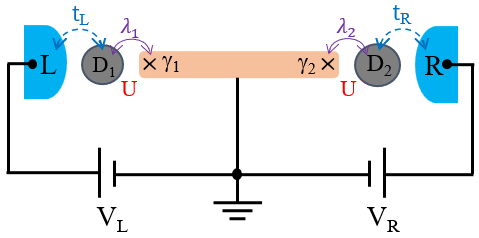} 
\caption{Schematic of the transport setup for a hybrid 
quantum dot-Majora zero mode (QD-MZM) system. 
Two QDs are coupled to a pair of Majorana modes ($\gamma_1$ and $\gamma_2$), 
located at the ends of a superconducting nanowire, with coupling strengths $\lambda_1$ and 
$\lambda_2$, respectively. Each QD is contacted by an electron
reservoirs via tunneling amplitudes $t_{\tL}$ and $t_{\tR}$. }
\label{fig1}
\end{figure}

The system under study, illustrated in \Fig{fig1},
consists of two quantum dots (QDs)
 coupled via a pair of Majorana zero
modes (MZMs), $\gamma_1$ and $\gamma_2$, located at the ends of 
a topological superconducting nanowire. Each QD is also connected 
to an individual normal electron reservoir.
The total Hamiltonian comprises three parts, $H_{\rm tot}=H_{\B}+H_{\s \B}+H_{\s}$.
The first part is the bath referring to the two electron reservoirs ($L$ and $R$),
which is given by $H_{\B}=\sum_{\alpha}\varepsilon_{\alpha k}
\hat c^\dg_{\alpha k}\hat c_{\alpha k}$, with $\alpha={ L,R}$.
%
The system-bath coupling Hamiltonian ($ H_{\s\B}$) describes the 
standard electrons tunneling between the QDs and electrodes:
  \begin{align}\label{Hcoup}
  H_{\s\B}&=  \sum_{\alpha k u}\left( t_{\alpha k u} \hat a^\dg_u \hat c_{\alpha k}+{\rm H.c.}\right),
 \end{align}
where $\hat a^\dg_u$ ($\hat a_u$) is the electron creation (annihilation) operator 
in the QD-$u$.
Throughout this work, we set $e=\hbar=1$, 
for the electron charge and
the Planck constant.


The central system is described by 
 $H_{\s}=H_0+H'+ H_{\U}$, which 
includes the bare Hamiltonian ($H_0$) of the dots and MZMs,
the QD-MZM coupling ($H'$), and Coulomb interaction between the electrons of the QDs and
the Majorana wire ($H_{\U}$). Within the low-energy effective description, 
these terms are given by
\cite{Lu12075318,Lu14195404,Fen21123032,Ric20165104},
\bsube\label{Htot0}
\begin{align}
  H_{0}&\!=\!\!\sum_{u=1,2}\!\varepsilon_u\hat a^\dg_u \hat a_u 
  \!+\! \frac{i}{2} \varepsilon_{\M}\hat\gamma_{1}\hat\gamma_{2}, 
  \label{H0}
\\
  H'&\!=\!
    \sum_{u=1,2}\! \lambda_{u}(\hat a^\dg_u-\hat a_u)\hat\gamma_{u},
 \\    
     H_{\U}&\!=\!   \sum_{u=1,2} U_u \hat n_{u}\hat n_{\M},
 \end{align}
 \esube
where each QDs is modeled by a single fermionic mode with creation 
(annihilation) operator $\hat a^\dg_u$ ($\hat a_u$)
 and energy level $\varepsilon_u$.
The coupling energy $\varepsilon_{\M}$ between the two MZMs
$\hat\gamma_1$ and $\hat\gamma_2$ depends on the wire length.
%
 The parameters $\lambda_{u}$ and $U_{u}$ denote, respectively,
 the coupling coefficients and the Coulomb interaction between the QD-$u$ and 
 the Majorana wire.
 Here,  $\hat n_u=\hat a^\dg_u \hat a_u$ and $\hat n_{\M}=\hat f^\dg \hat f$ 
 are the occupation operators of the dots and wires, respectively.
%
%

The Majorana operators can be expressed in terms of regular fermionic operators via
  $\hat\gamma_{1}=\hat f +\hat  f^\dg $ and $\hat\gamma_{2} = -i(\hat f -\hat  f^\dg)$,
 where $\hat  f^\dg$ ($\hat f $) is the fermionic creation (annihilation) operator.
 In this representation, 
the Hamiltonian \Eq{Htot0} becomes
\bsube\label{Htot}
\begin{align}
  H_{0}&\!=\!\sum_{u=1,2}\varepsilon_u\hat a^\dg_u \hat a_u
  + (\varepsilon_{\M}\!-\!\half)\hat f^\dg \hat f,
\\
  H'&\!=\!
    \sum_{u=1,2}\! \lambda_u\left[\hat a^\dg_u\hat f+(-)^u\hat a_u^\dg\hat f^\dg+{\rm H.c.}\right],
 \\    
     H_{\U}&\!=\!   
     \sum_{u=1,2} \!\half U_u \hat n_u(i\hat\gamma_1\hat\gamma_2+1).
 \end{align}
 \esube
 In this form,
the coupling $H'$ between QDs and MZMs consists of two distinct processes:
the normal tunneling (NT) term, given by 
$\lambda[\hat a^\dg_u \hat f \!+\! {\rm H.c.}]$,
and anomalous tunneling (AT) term, given by 
$\lambda[\hat a^\dg_u \hat f^\dg \!+ \!{\rm H.c.}]$.
 Notably, the structure of $H_{\U}$ closely resembles the term
 $ \varepsilon_{\M}\hat\gamma_{1}\hat\gamma_{2}$ in $H_0$ \Eq{H0}. 
 Therefore, the
QD-MZM Coulomb interaction leads to additional hybridization of the
Majorana modes, whose amplitude $\!\half U_u \hat n_u$ depends on occupancy
of the QD-$u$. As shown below, this
 has a dramatic effect on Majorana-mediated electron teleportation between the two QDs.

 
%

\subsection{The DEOM approach}

In this work, we employ the established 
  fermionic dissipaton equation of emotion (DEOM) method for
   accurate evaluations on the transport current
  and its noise spectrum \cite{Yan14054105,Jin15234108,Yan16110306}.
 The core idea of DEOM is the introduction of the {\it dissipatons}, quasi–particles
 used to decompose the bath operators: 
\be\label{wtiF_f}
  \wti F^{\sigma}_{\alpha u} \equiv -\sigma \hat  F^{\sigma}_{\alpha u}
    \equiv \sum_{m=1}^{M} \hat f^{\sigma}_{\alpha u m},
\ee
where $ \hat F^-_{\alpha u}
=\sum_k t_{\alpha u k} c_{\alpha k}=(\hat F^+_{\alpha u})^\dg$ and
 $\sigma =+,-$ (with $\bar\sigma$ denoting the opposite sign) 
identifies the creation/annihilation operators.  
The system-reservoir coupling in \Eq{Hcoup} can then be expressed as
\be\label{Hsb1}
  H_{\SB}\!=\!\sum_{\alpha u \sigma}\hat a^{\bar\sigma}_{u}
    \wti F^{\sigma}_{\alpha u}\!=\!\sum_{\alpha u \sigma m}\hat a^{\bar\sigma}_{u}
    \hat f^{\sigma}_{\alpha u m}.
\ee
Within the fluctuation-dissipation framework, the bath correlation function
 $\big\la \hat F^{\sigma}_{\alpha u}(t)
\hat F^{\bar\sigma}_{\alpha v}(0)\big\ra_{\B}=\frac{1}{\pi}\int_{-\infty}^{\infty}d\omega e^{\sigma i\omega t}
 J^\sigma_{\alpha uv}(\omega)f^\sigma_\alpha(\omega)$ is considered for
 a Lorentzian hybridization spectral function,
\be\label{Jw}
J_{\alpha uv}(\omega)
\equiv\pi\sum_k t_{\alpha  k}t^\ast_{\alpha  k}\delta(\omega-\varepsilon_{\alpha k})
=\frac{\Gamma_{\alpha uv}W^2}{\omega^2+W^2}.
\ee
Here $J_{\alpha uv}(\omega)=J^{-}_{\alpha vu}(\omega)=J^+_{\alpha vu}(\omega)$, 
and $f^+_\alpha(\omega)=1-f^-_\alpha(\omega)$ is the Fermi distribution function. 
 %
%
Using an optimal Pad\'{e} spectrum decomposition
of the Fermi function \cite{Hu10101106,Hu11244106},
the bath correlation function can be written in exponential
 form, 
\be\label{FF_corr}
\big\la \hat F^{\sigma}_{\alpha u}(t)
\hat F^{\bar\sigma}_{\alpha v}(0)\big\ra_{\B}
 \!=\! \sum_{m=1}^{M} \eta^{\sigma}_{\alpha uv m}e^{-\gamma^{\sigma}_{\alpha m}t},
\ee
which leads to
\be\label{ff_corr}
   \big\la\hat f^{\sigma}_{\alpha u m}(t)\hat f^{\sigma'}_{\alpha' v m'}(0)\big\ra_{\B}
 =-\delta_{\sigma\bar\sigma'}\delta_{\alpha\alpha'}\delta_{mm'}\,
   \eta^{\sigma}_{\alpha u v m}\, e^{-\gamma^{\sigma}_{\alpha m} t}.
   \nonumber
\ee
 Introducing the composite index
 $j\equiv(\sigma\alpha u m)$ and $\bar j\equiv(\bar\sigma\alpha u m)$,
the dynamical variables in DEOM are the reduced dissipaton density
operators (DDOs):
\be\label{DDO_def}
 \rho^{(n)}_{\bf j}(t)\equiv \rho^{(n)}_{j_1\cdots j_n}(t)\equiv
 {\rm tr}_{\B}\Big[\big(\hat f_{j_n}\cdots\hat f_{j_1}\big)^{\circ}
  \rho_{\rm tot}(t)\Big]\, ,
\ee
where the product inside the circled $(\,\cdot\cdot\,)^{\circ}$
is \emph{irreducible}.

 From the perspective of quantum dissipation theory, 
  the central QD-MZM hybrid system is described by the
reduced system density operator $\rho(t)\equiv{\rm tr}_{\B}[\rho_{\rm tot }(t)]$.
  Taking the time derivative and tracing over the reservoirs leads to the 
   the DEOM (equivalent to the hierarchical equations of motion, HEOM)
 \cite{Yan14054105,Jin15234108,Yan16110306},
\begin{align}\label{DEOM}
  \dot\rho^{(n)}_{\bf j}(t)&=-\bigg(i{\cal L}_{\tS}
  +\sum_{r=1}^n \gamma_{j_r}\bigg)\rho^{(n)}_{\bf j}(t)
  -i\sum_{j} {\cal A}_{\bar j}\rho^{(n+1)}_{{\bf j}j}(t)    \nl
&\quad
  -i \sum_{r=1}^n (-)^{n-r}{\cal C}_{j_r}\rho^{(n-1)}_{{\bf j}^-_r}(t),
\end{align}
where,
 ${\cal L}_{\tS}\,(\cdot)=[H_{\tS},\,(\cdot)]$
and the superoperators, ${\cal A}_{\bar j}\equiv {\cal A}^{\bar\sigma}_{\alpha u m} = {\cal A}^{\bar\sigma}_{u}$
and ${\cal C}_{j}\equiv {\cal C}^{\sigma}_{\alpha u m}$
 are defined via 
\be\label{calAC}
\begin{split}
 {\cal A}^{\sigma}_{u} \Opm &\equiv
    a^{\sigma}_{ u}\Opm \pm \Opm \hat a^{\sigma}_{u}
 \equiv \big[\hat  a^{\sigma}_{u},\Opm\big]_\pm \, ,
\\
 {\cal C}^{\sigma}_{\alpha u m} \Opm  &\equiv
  \sum_{v} \big(\eta^{\sigma}_{\alpha uv m}\hat  a^{\sigma}_{v}\Opm
  \mp \eta^{\bar \sigma\,{\ast}}_{\alpha uv m}\Opm \hat a^{\sigma}_{m}\big),
\end{split}
\ee
with $\Opm$ denoting an arbitrary operator of
 even ($+$) or odd ($-$) fermionic parity.
%
The reduced system density operator is
$\rho(t) \equiv {\rm tr}_{\B}\rho_{\rm tot}(t)= \rho^{(0)}(t)$.
 All higher-tier DDOs $\big\{\rho^{(n\geq1)}_{\bf j}\big\}$
captures entangled system--bath dynamics and are physically well-defined in \Eq{DDO_def}.

The DEOM provides a unified framework for both noninteracting 
and interacting open quantum systems. 
For noninteracting systems, the hierarchy terminates exactly at tier
 $L=2$. 
  In fact, truncating at  $L=2$ 
  yields an exact equation for the
  single-particle density matrix,  
  $\varrho_{uv}={\rm tr}_{\s}[a^\dg_v a_u\rho(t)]$,
    and exact
  current expression \cite{Jin08234703},
consistent with the Landauer–Büttiker scattering theory
  \cite{Dat95} and nonequilibrium Green function method \cite{Hau08}.
For interacting systems,
DEOM converges uniformly and rapidly as the truncated tier $L$ (where $\rho^{(n>L)}_{\bf j}=0$) 
increases, with the required $L$ 
 depending on the specific system-bath configurations \cite{Zhe121129,Li12266403,%
Zhe13086601,Hou15104112,Ye16608}.
%

 Equation (\ref{DEOM}) constitutes a set of linear
differential equations, enabling the use of standard quantum dynamics pictures in
the DEOM-space descriptions of open quantum systems \cite{Yan16110306}.
This structure supports consistent
DEOM-based evaluations of various physical quantities, including transport current
and its correlation function, as outlined below.

 \subsection{ Transport current and correlation function}
The transient current though the lead-$\alpha$ is calculated via
$I_{\alpha}(t) = {\rm Tr}[\hat I_{\alpha}\rho_{\rm tot}(t)]$, with the current operator
 $ \hat I_{\alpha} \equiv -\dot{\hat N}_{\alpha}=-i\sum_u \big(\hat  a^{+}_u \hat F^-_{\alpha u}
   -\hat F^{+}_{\alpha u}\hat  a^-_u \big)$. This leads to
\be\label{Acurr}
  I_{\alpha}(t) = -i\! \sum_{j_{\alpha}\in j}
  {\rm tr}_{\tS}\!\big[\ti a_{\bar j}\rho^{(1)}_{j}(t)\big],
\ee
where $\ti a_{\bar j}\equiv \ti a^{\bar \sigma}_{\alpha u k}
=\bar \sigma\hat a^{\bar \sigma}_{u}$
and $j_{\alpha}\equiv \{ \sigma u k\}\in j\equiv\{\sigma\alpha u k\}$.
The current is therefore directly related to the first-tier DDOs.

Similarly,the current-current correlation function
can be expressed as 
\begin{align}\label{curr-curr}
 \La\hat I_{\alpha}(t)\hat I_{\alpha'}(0)\Ra
&= 
{\rm Tr}\big[\hat I_{\alpha}\rho_{\rm tot}(t;\hat I_{\alpha'})\big]
\nl&=- i\!\sum_{j_{\alpha}\in j}
{\rm tr}_{\tS} \big[\ti a_{\bar j}
 {\rho}^{(1)}_u(t;\hat I_{\alpha'})\big],
\end{align}
which also depends on the first-tier DDOs, but with a different
 initial condition: $\rho_{\rm tot}(0;\alpha')=\hat I_{\alpha'}\rho^{\rm st}_{\rm tot}$,
where $\rho^{\rm st}_{\rm tot}$ is the steady state of the total system. 
Using dissipaton-algebra, the initial
DDOs are given by 
\begin{align}\label{initial_final}
 \rho^{(n)}_{\bm j}(0; \alpha')
&\!=\! -i\!\!\!\sum_{j'_{\alpha'}\in j'}{\ti a}_{\bar j'} \rho^{(n+1);{\rm st}}_{{\bm j}j'}
 - i\!\sum_{r=1}^n(-)^{n-r} \wti C_{j_r}\rho^{(n-1);{\rm st}}_{{\bm j}^{-}_r}.
 \nonumber
\end{align}
Further details can be found in Refs. \cite{Yan14054105,Jin15234108,Yan16110306}
and numerical applications \cite{Jin20235144,Mao21014104}.

%
The steady--state mean current is denoted by $\bar I_{\alpha}
=  \la \hat I_{\alpha}\ra$
 and its fluctuation is given by,
\be\label{corr-curr}
  \La \delta{\hat I}_\alpha(t)\delta{\hat I}_{\alpha'}(0)\Ra
=\La [{\hat I}_\alpha(t)-\bar I_{\alpha}]
  [{\hat I}_{\alpha'}(0)-\bar I_{\alpha'}]\Ra.
  \nonumber
\ee
The lead-specified noise spectrum is the Fourier transform,
\be\label{Sw_alp}
  S_{\alpha\alpha'}(\omega)=\int_{-\infty}^{\infty} \!dt\,
  e^{i\omega t} \La \delta{\hat I}_\alpha(t)\delta{\hat I}_{\alpha'}(0)\Ra.
\ee
Unlike the symmetrized version
$S^{{\rm sym}}_{\alpha\alpha'}(\omega)
=S_{\alpha\alpha'}(\omega)
+S_{\alpha'\alpha}(-\omega)$,
the asymmetric $S_{\alpha\alpha'}(\omega)$
directly relates to absorption ($\omega>0$) and ($\omega<0$) emission processes
\cite{Eng04136602,Rot09075307,Yan14115411,
Bas10166801,Bas12046802,Del18041412}.
In this work, we focus on 
the cross-correlation spectrum ${\rm Re}[{S_{\tL\tR}(\omega)}]$
and its zero-frequency value $S_{\tL\tR}(0)=S_{\tR\tL}(0)$.

\section{Isolated hybrized QD-MZM system}
\label{thham}

In the Fock states basis $\{|n_{1}n_{\M}n_2\ra \}$,
the Hamiltonian in \Eq{Htot} becomes block diagonal. It separates into
subspaces of odd parity (${ p}= -1$), spanned by $\{|100\ra, |010\ra, |001\ra,|111\ra \}$, 
and even parity (${ p}= +1$), spanned by $\{|110\ra,|000\ra, |011\ra,|101\ra\}$:
\begin{align}\label{Hsm}
H_{\s}={
  \left( \begin{array}{cc}
           H_{-} & 0 \\
           0 & H_{+}
         \end{array}
         \right)},
\end{align}
where the sub-block Hamiltonians are given by
\bsube\label{Hpn}
\begin{align}\label{Hodd}
H_{-}&\!=\!{
\left( \begin{array}{cccc}
         \varepsilon_{1} & \lambda_1 & 0 & \lambda_2 \\
       \lambda_1 & \varepsilon_{\M} & \lambda_2 & 0 \\
       0 & \lambda_2 & \varepsilon_2 & \lambda_1 \\
       \lambda_2 & 0 & \lambda_1 & \varepsilon_1+\varepsilon_2+2 U
       \end{array}
       \right)},
\end{align}
\begin{align}\label{Heven}
H_{+}&\!\!=\!\!{
\left( \begin{array}{cccc}
         \varepsilon_{1}+\varepsilon_{\M}+U & \lambda_1 & 0 & \lambda_2 \\
       \lambda_1 & 0 & \lambda_2 & 0 \\
       0 & \lambda_2 & \varepsilon_2+\varepsilon_{\M}+U & \lambda_1 \\
       \lambda_2 & 0 & \lambda_1 & \varepsilon_1+\varepsilon_2
       \end{array}
       \right)}.
\end{align}
\esube
Here, we assume the coupling coefficients ($\lambda_1$ and $\lambda_2$) between 
the QDs and the MZMs are real, and we set $U_1=U_2=U$.
In this description of the isolated hybrid QD-MZM system,
  the parity of $H_{\s}$ is a conserved quantity, since no
  transitions occur between the two parity subspaces.
  %
 The focus of the present work is to investigate the effect of Coulomb interaction $U$
  on nonlocal behavior in the limit of zero Majorana coupling energy
 $\varepsilon_{\M}\rightarrow0$. Therefore, we set $\varepsilon_{\M}=0$ 
 in the following discussions unless stated otherwise.

\begin{figure}
\includegraphics[width=1.\columnwidth]{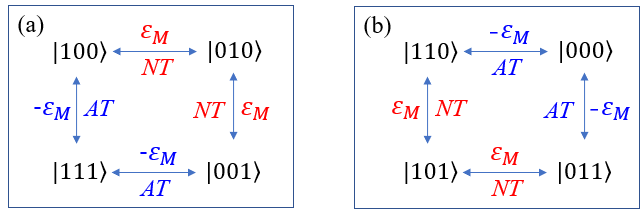} 
\caption{ Coherent electron transitions between the states 
within the odd-parity subspace (a)
and the even-parity subspace (b).
The dynamics in each subspace are governed 
by the Hamiltonian $H_-$ [\Eq{Hodd}] 
and $H_+$ [\Eq{Heven}], respectively.
Both involve two distinct tunneling process:  
 normal tunneling (NT) via the electron channel $\varepsilon_{\M}$
 and anomalous tunneling (AT)
 via the hole channel $-\varepsilon_{\M}$. }
\label{fig2}
\end{figure}
\subsection{Noninteracting case: $U=0$}

We first consider the noninteracting case with $U=0$,
where the Hamiltonian satisfies $H_- = H_{+}$.
Starting from the state $|100\ra$ (or $|110\ra$), the quantum evolution 
is confined to the odd-parity (or even-parity) subspace.
The time-dependent occupation probabilities of the electron 
in the left and the right dots are given by
%
\bsube
\begin{align}\label{odd0}
  P_1(t)&=1 -\frac{4\lambda_1^2}{\Lambda_1^2} \sin^2(\!\!\half \Lambda_1^2 t),
 \\
   P_2(t)&=\frac{4\lambda_2^2}{\Lambda_2^2} \sin^2(\!\!\half \Lambda_2 t),
\end{align}
\esube
where 
  $\Lambda_1^2\equiv\varepsilon_1^2 + 4\lambda_1^2$  and 
  $\Lambda_2^2\equiv\varepsilon_2^2 + 4\lambda_2^2$.
 Consistent with previous studies \cite{Tew08027001,Fen21123032},
  the occupation probabilities of the two dots are independent. For instance,
the occupation probability of the right dot
does not depend on the coupling strength $\lambda_1$, the energy $\varepsilon_1$, and
the occupation of the left dot.
This implies that an electron initially in the left dot cannot be teleported to the right dot---
a consequence of destructive interference between the two tunneling channels 
mediated by the MZMs.
  %

Specifically, $P_2(t)$ can be expressed as the sum of two probabilities, 
 for odd-parity subspace, $P_2(t)=P_{001}(t)+P_{111}(t)$, and for even-parity subspace, 
 and $P_2(t)=P_{011}(t)+P_{101}(t)$. These components are:
\bsube
 \begin{align}\label{p1u0}
 P_{001}(t)&\!=\!P_{011}(t)\!=\!\frac{4\lambda_1^2 \lambda_2^2}{\Lambda_1^2 \Lambda_2^2} 
 \big[ \cos (\!\!\!\half \wti\Lambda t )
 \!-\! \cos (\!\!\half  \bar\Lambda t ) \big]^2,
\\
P_{111}(t)&\!=\!P_{101}(t)\!= \!\frac{4\lambda_2^2}{\Lambda_2^2} 
  \sin^2 (\!\!\half  \Lambda_2 t) 
\nl &
\quad\quad\quad
- \frac{4\lambda_1^2 \lambda_2^2}{\Lambda_1^2 \Lambda_2^2}  \big[ \cos(\!\!\half \wti\Lambda t )
- \cos(\!\!\half \bar\Lambda t )\big]^2,
\label{p2u0}
\end{align}
\esube
  where $ \wti\Lambda\equiv\Lambda_1 +\Lambda_2 $ and $ \bar\Lambda\equiv\Lambda_1 -\Lambda_2 $.
The probabilities in \Eq{p1u0} and \Eq{p2u0} 
 arise from the two coherent tunneling channels: 
  normal tunneling 
 (via the electron channel $\varepsilon_{\M}$) and anomalous tunneling
 (via the hole channel $-\varepsilon_{\M}$), as illustrated in \Fig{fig2}. 
 These two channels become 
 degeneracy when $\varepsilon_{\M}=0$, leading to 
  quantum destructive interference.
 %
 In this degenerate limit, the processes contributing to
  $P_{001}$ and $P_{111}$ (for the odd-parity example)
  are indistinguishable, resulting in independent
 dot occupations and vanishing cross-correlations. 
  %
  Lifting this degeneracy makes the combined probabilities $P_{001}+P_{111}$ 
  depend explicitly on $\lambda_1$.
  Consequently, the occupations of the two dots become correlated, giving rise to
   nonzero cross correlations 
  and enabling electron transfer between the dots via
 spatially separated Majorana modes.
  This teleportation effect was originally proposed in Ref.\cite{Fu10056402},
by taking the large charging energy limit ($E_C\rightarrow\infty$)
 on the Majorana island, which fully suppresses the 
 AT channel and isolates the NT channel.
  %
  %
  More recently, Feng {\it et al.,} \cite{Fen21123032} demonstrated the effect of 
  finite charging energy $E_C$ to break the degeneracy, 
  yielding nonzero cross correlation in the limit   
$\varepsilon_{\M}\rightarrow0$. 
  In the present work, we propose an alternative scheme to remove the degeneracy:
introducing finite Coulomb interaction between QDs and the Majorana wire.
  %


  \begin{figure}
\centering
\centerline{\includegraphics*[width=0.75\columnwidth,angle=0]{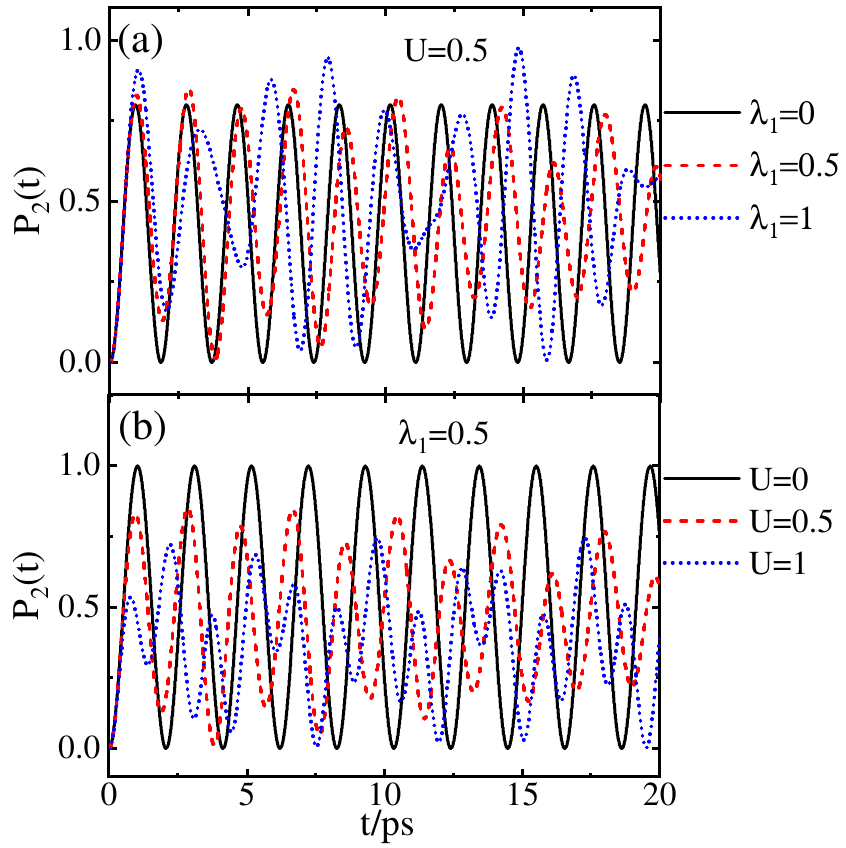}}
 \caption{ The time evolution of the occupation probability in the right dot $P_2(t)$
 for (a) various $\lambda_1$ with $U=0.5$ and (b) various $U$ for $\lambda_1=0.5$.
 The initial state is $|110\ra$ with $P_{110}(0)=1$.
  \label{fig3}}
\end{figure}

\subsection{Interacting case: $U\neq 0$}
When Coulomb interaction $U$ is present, 
the Hamiltonians of odd- and even-parity subspaces are no longer equivalent,
i.e., $H_- \neq H_{+}$, as seen in \Eq{Hpn}.
In the even-parity subspace described by \Eq{Heven},
the parameters $U$ and $\varepsilon_{\M}$ play similar roles.
In contrast, their roles differ in the odd-parity Hamiltonian
   $H_-$ [\Eq{Hodd}]: transitions between
 $|111\ra$ and $|100\ra$ (or $|001\ra$) requires 
 an additional energy $2U$ when proceeding through
  the AT channel compared to the NT channel, see \Fig{fig2} (a).
Despite this distinction, the underlying physical mechanism 
is the same---the degeneracy between the two tunneling channels is lifted.
This breaking of degeneracy can enable electron teleportation, 
allowing an electron initially in the left dot can transfer to the right dot via the 
spatially separated Majorana modes. 
 For clarity, without loss of generality, we further set 
  the energy levels in the dot as $\varepsilon_1=\varepsilon_2=0$ in the following description.

For the even-parity subspace governed by \Eq{Heven}, 
starting from the initial state $|110\ra$, 
the time-dependent occupation probabilities of the electron in the left dot and right dot
are given by $P_1(t)=P_{110}(t)+P_{101}(t)$ and $P_2(t)=P_{011}(t)+P_{101}(t)$, respectively.
Explicitly, 
\bsube \label{even-p12t}
\begin{align}
\label{even-p1t}
  P_1(t)&
  = \frac{1}{2}\Big[1+ \cos\big(\half \bar Ut \big)
 \cos\big( {\half \wti{U}t} \big) 
 \nl&\quad\quad
 -\frac{4\bar\lambda\ti\lambda-U^2}{\wti U\bar U}
 \sin\big( {\half \bar{U}t} \big)
 \sin\big( {\half \wti{U}t} \big)\Big],
  \\
  P_2(t)&=\frac{1}{2}\Big[1- \cos\big(\half \bar Ut \big)
 \cos\big( {\half \wti{U}t} \big)
 \nl&\quad\quad
  -\frac{4\bar\lambda\ti\lambda+U^2}{\wti U\bar U}
 \sin\big( {\half \bar{U}t} \big)
 \sin\big( {\half \wti{U}t} \big)\Big],
 \label{even-p2t}
\end{align}
\esube
where 
\be\label{tiu}
\wti U\equiv \sqrt{U^2+4\ti\lambda^2},~~~\bar{ U}\equiv \sqrt{U^2+4\bar\lambda^2},
\ee
with $\ti\lambda=\lambda_1+\lambda_2$ and $\bar\lambda=\lambda_1-\lambda_2$.  
 Clearly,  
 the charge occupations of the two QDs described by \Eq{even-p12t}
are mutually dependent.  
We note that the same expression 
can be obtained by considering a finite
$\varepsilon_{\M}$ and replacing $U$ with $\varepsilon_{\M}$ in \Eq{even-p12t},
since the two parameters enter the Hamiltonian $H_+$ in an analogous manner.

For the odd-parity subspace
described by \Eq{Hodd}, 
 analytical solutions are not readily obtainalbe.
 We therefore present the numerical results, 
as shown in \Fig{fig3}.
Figure (3a) illustrates that $P_2(t)$ becomes sensitive to $\lambda_1$ when $U$ is finite.
Moreover, the interaction $U$ modifies both the amplitude and the oscillation frequency of $P_2(t)$
as seen in \Fig{fig3} (b).
These modifications become more pronounced with increasing $U$,
indicating that
the nonlocal correlation between the two QDs becomes 
strengthens as Coulomb interaction grows.
Throughout this paper, we set $\lambda_2=1 ({\rm meV})$ as the unit of the
 energy. 


\section{Quantum measurement}
\label{thmea}
To probe this nonlocal teleportation effect, we examine quantum transport through
the QD-MZM hybrid system, as illustrated in \Fig{fig1}.
%
The transport current or conductance can reflect the probability information.
Alternatively, one may measure the current noise spectrum--particularly 
the cross-correlation noise, 
including its zero-frequency and the finite-frequency components.
It is worth noting that, within this transport measurement setup, 
parity is no longer conserved.
Taking again the initial state $|100\ra$ from the odd-parity subspace as an example, 
states across the entire Hilbert space--spanning both odd- and even-parity subspaces--
 will participate in the time evolution
 due to the tunneling couplings with the two electrodes.
After tracing out the freedom of the two electron reservoirs,
the QD-MZM hybrid system is described by the
reduced system density operator $\rho(t)\equiv{\rm tr}_{\B}[\rho_{\rm tot }(t)]$.
The exact dynamics of $\rho(t)$ follows the DEOM given by \Eq{DEOM}.
%

In the following numerical calculations, we consider the symmetrical transport configurations 
with $\Gamma_{\tL}=\Gamma_{\tR}=\Gamma/2$ and $\mu_{\tL}=\mu_{\tR}=V/2$,
unless specified otherwise. 
We set $\Gamma=k_{\B}T=0.1$ and $W=100$ for wide band reservoirs.

  \begin{figure}
\centering
\centerline{\includegraphics*[width=0.95\columnwidth,angle=0]{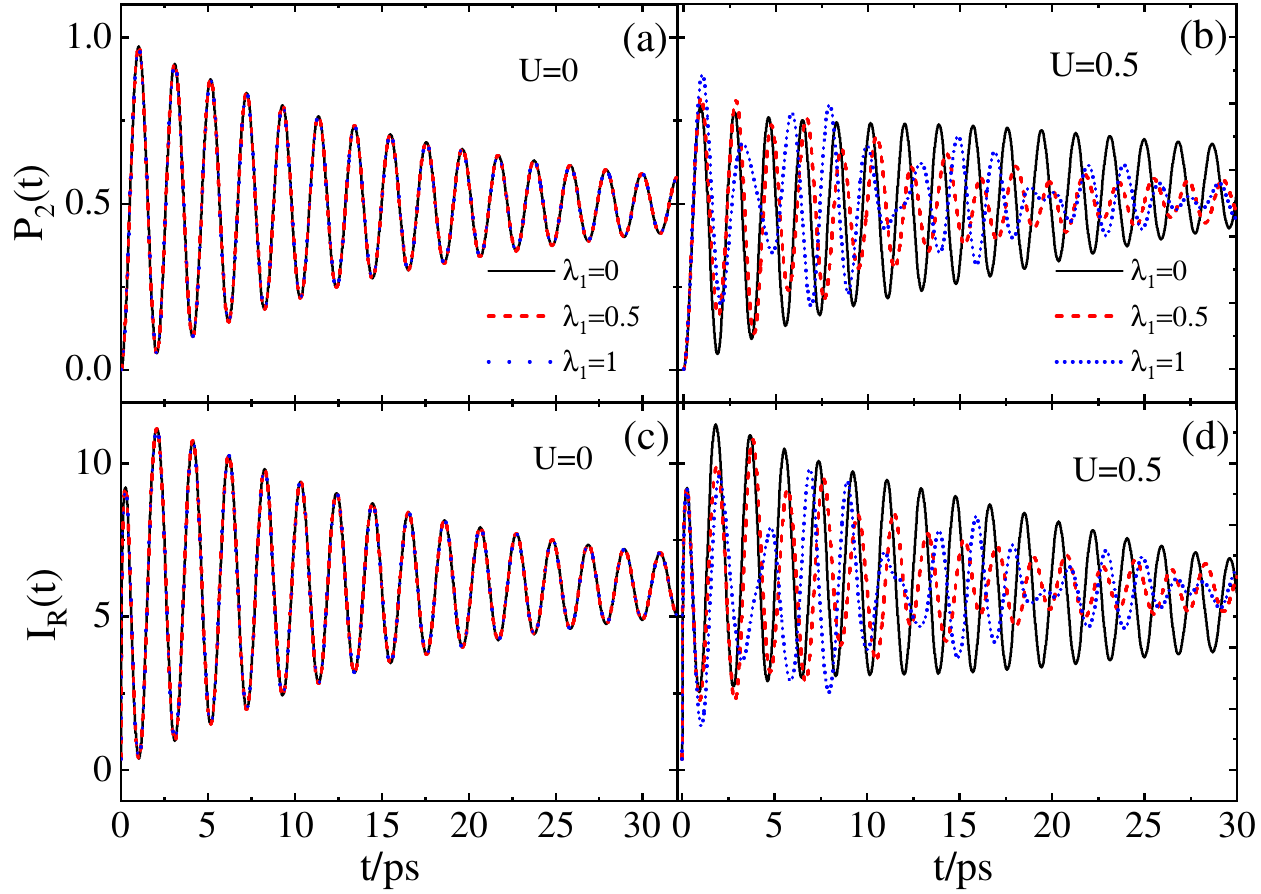}}
 \caption{ The time evolution of the occupation probability 
 in the right dot $P_2(t)$,
 and the transport current through the right electrode $I_R(t)$.
 (a) For viarious $\lambda_1$ with $U=0.5$;  (b) For various $U$ for $\lambda_1=0.5$.
 The initial state is $|110\ra$ with $P_{110}(0)=1$. 
  \label{fig4}}
\end{figure}
\subsection{The transport current} 
We examine the transient current in the right electrode $I_{\tR}(t)$,
using the initial state $|100\ra$. The numerical results are presented in \Fig{fig4},
where the time evolution of the average current $I_{\tR}(t)$ closely follows 
that of the dot occupation probability $P_2(t)$.

In the absence of the Coulomb interaction ($U=0$), 
both $I_{\tR}(t)$ and $P_2(t)$ are insensitive to
 the coupling strength $\lambda_1$, as shown in 
the left column of \Fig{fig4}. This
 indicates that the transient currents in the two electrodes [$I_{\tL}(t)$
 and $I_{\tR}(t)$] are 
 independent each other, 
 reflecting the independent 
 occupation probabilities of the right and left QDs.
In contrast, when Coulomb interaction is present ($U\neq0$),
  both $I_{\tR}(t)$ and $P_2(t)$ become distinctly sensitive to
  $\lambda_1$--a signature of electron teleportation--
  as displayed in the right column
of \Fig{fig4}.
These characteristics are fundamentally consistent with 
the features of the occupation probabilities
in the isolated system (without electrodes), discussed in \Sec{thham}
and shown in \Fig{fig3}.
The primary effect of coupling to the electrodes is to
 introduce a damped oscillation decay in 
 the occupation dynamics. 
In a word, the transport current effectively 
captures the teleportation effect. 
%

Although only the transient current is displayed here, 
the same behavior applies to the steady-state
current, differential conductance, 
and auto-correlation noise spectrum.
All these quantities can reveal the teleportation phenomena,
as they share the same $\lambda_1$-dependence (or lack thereof) 
in the respective cases 
with and without the Coulomb interaction.
To further explore this nonlocal characteristic, we next
investigate the cross-correlation noise spectrum, which provides a more
direct and efficient measure of  
 nonlocal correlation, as demonstrated in the following subsection.


%
\subsection{Cross-correlation noise spectrum}
\label{thnoise}

\begin{figure}
\centering
\centerline{\includegraphics*[width=1.0\columnwidth,angle=0]{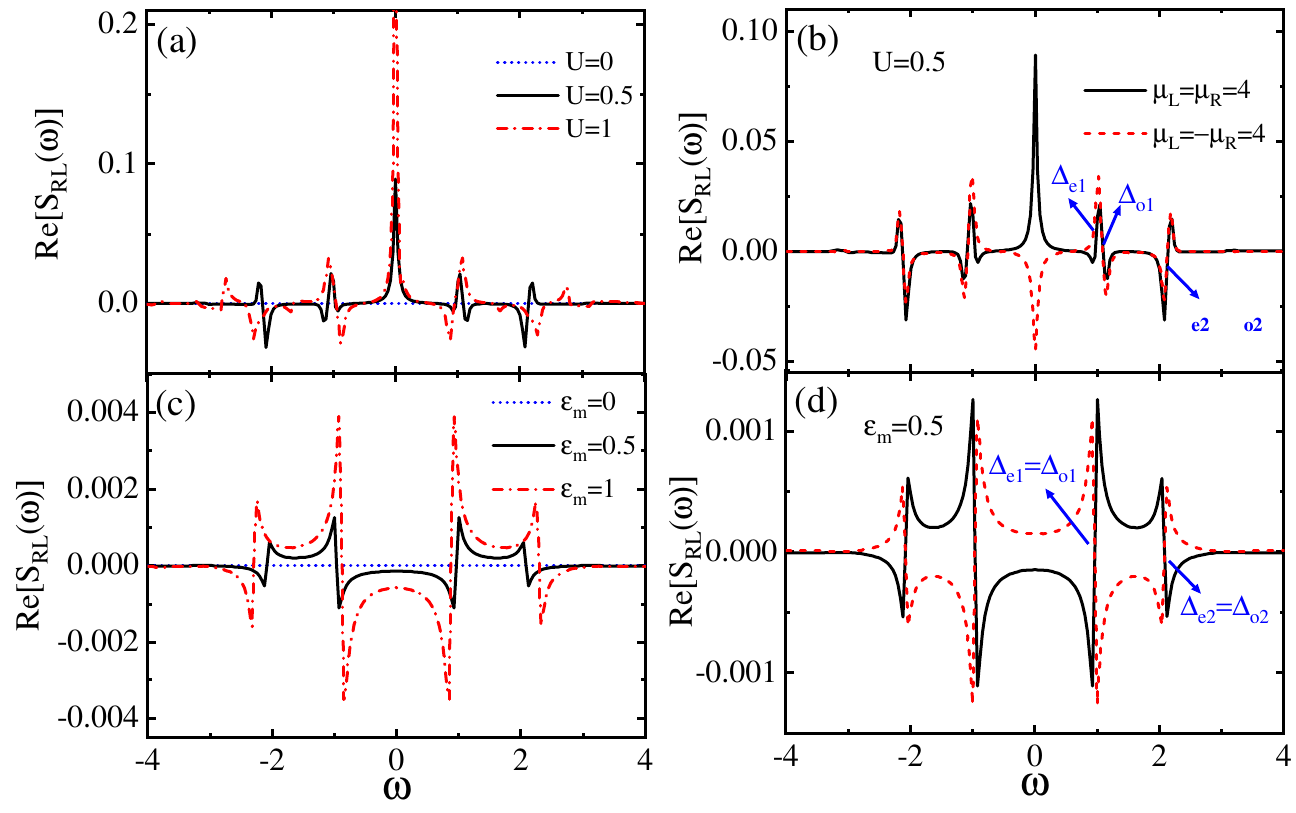}}
 \caption{ The cross-correlation noise spectrum ${\rm Re}[{S_{\tL\tR}(\omega)}]$ (in units of $e\lambda_2/\hbar$).
 In the up panels, we consider $\varepsilon_{\M}=0$ for (a) different Coulomb interaction $U$ with $\mu_{\tL}=\mu_{\tR}=4$, 
 and (b) different chemical potential configurations with $U=0.5$.
 For comparison, we also plot the result of $U=0$ but finite $\varepsilon_{\M}$
 in the low panels:
 for (c) different $\varepsilon_{\M}$ values, and 
 (d) different chemical potential configurations with $\varepsilon_{\M}=0.5$.
   \label{fig5}}
\end{figure}

 We now examine the 
current cross-correlation noise spectrum, a widely used tool for  
probing the non-local behavior \cite{Law09237001,Liu13064509,
  Ulr15075443,Bol07237002,Nil08120403,Zoc13036802, 
  Bjo13036802,Lu14195404,Lu16245418, Fen21123032}.
Our accurate numerical results are presented in \Fig{fig5}.
For $U=0$, ${\rm Re}[{S_{\tL\tR}(\omega)}]\rightarrow 0$, 
indicating no correlation between the two QDs.
In contrast, for $U\neq 0$, ${\rm Re}[{S_{\tL\tR}(\omega)}]$ exhibits rich structures.
Its nonzero value directly
reflects the electron teleportation between the two QDs via the Majorana wire.
Moreover, the peak-dip features in ${\rm Re}[{S_{\tL\tR}(\omega)}]$
reveal coherent dynamics between the QDs and Majonara wires within
 each parity subspace.
These features occur at characteristic frequencies
determined by the energy difference between eigenstates.

For instance, in the even-parity subspace described by \Eq{Heven},
the eigenenergies are,
\begin{align}\label{eigen}
  E_1&=\frac{1}{2}(U-\bar U),~~~
  E_2=\frac{1}{2}(U+\bar U),
  \nl
  E_3&=\frac{1}{2}(U-\wti U),~~~
  E_4=\frac{1}{2}(U+\wti U),
\end{align}
with $\wti U$ and $\bar U$ given by \Eq{tiu}.
Peak-dip features appear at frequencies 
$\omega\simeq \pm\Delta_{e1}$ and $\omega\simeq \pm\Delta_{e2}$, where
\bsube
\begin{align}\label{eigen-w}
\Delta_{e1}&=|E_1-E_3|=|E_2-E_4|=\wti U-\bar U,
\\
\Delta_{e2}&=|E_1-E_4=|E_2-E_3|=\wti U+\bar U.
\end{align}
\esube
Although analytical expressions for the eigenenergies in the odd-parity subspace 
are not readily obtainable, both parity sector exhibit similar characteristic frequencies,
 as confirmed numerically.
 For the parameters in \Fig{fig5} (b) with $U=0.5$, 
 we find $\Delta_{e1}=0.96$, $\Delta_{o1}\approx1.14$ and 
 $\Delta_{e2}\sim\Delta_{o2}\approx2.1$ (indicated by solid arrows). 
 In generally, the correlation signal in ${\rm Re}[{S_{\tL\tR}(\omega)}]$
 increases with Coulomb interaction, 
 particularly near zero-frequency.
 As shown in \Fig{fig5} (a), ${\rm Re}[{S_{\tL\tR}(0)}]\approx0.4$
 for $U=1$.

Furthermore, the peak-dip profiles resemble Fano line shapes,
 indicating interference between the NT and AT channels.
  If only NT or AT is present,
the profiles would reduce to simple peaks or dips. 
 These characteristics are analogous to those observed in double-dot Aharonov-Bohm 
 interferometer \cite{Jin20235144}.
 Interestingly, the Fano profiles are only weakly affected by bias voltage,
 as seen in \Fig{fig5} (b) when comparing symmetrical ($\mu_{\tL}=\mu_{\tR}$) 
 and antisymmetrical ($\mu_{\tL}=-\mu_{\tR}$) bais configurations.
However, the zero-frequency response changes dramatically,
 switching from the positive to negative. 
 %
 %
 Additionally,  ${\rm Re}[{S_{\tL\tR}(0)}]$ is larger in the
 symmetrical case, suggesting that the nonlocal 
 correlation are more easily observed in the zero-frequency 
 component under symmetric biasing.

For comparison, we also examine the effect of the finite Majorana coupling energy $\varepsilon_{\M}$ 
on non-lcoal behavior, as discussed in Ref.\cite{Lu14195404} 
for the zero-frequency noise [${S_{\tL\tR}(0)}$].
Here, we plot the frequency-dependent correlation noise ${\rm Re}[{S_{\tL\tR}(\omega)}]$
for finite $\varepsilon_{\M}$ and $U=0$ 
in the lower panels of \Fig{fig5} [(c) and (d)].
In this case, $H_-=H_+$ for $\varepsilon_1=\varepsilon_2=0$ [\Eq{Hsm}].
The degenerate NT and AT channels split symmetrically around $\varepsilon_{\M}$ and  
 $-\varepsilon_{\M}$, respectively (see \Fig{fig2}).
The resulting ${\rm Re}[{S_{\tL\tR}(\omega)}]$ 
displays dip-peak Fano profiles at
the eigenfrequencies $\omega\simeq \pm\Delta_{e1} =\pm \Delta_{o1}$ 
and $\omega\simeq \pm\Delta_{e2}=\pm\Delta_{o2}$ (solid-arrows in \Fig{fig5} (d)]. 
Upon reversing the bias from symmetric to antisymmetric,
the correlated noise exhibits a mirror-image behavior.

Comparing the correlation signal  
 induced by finite Coulomb interaction $U$ [upper panels (a) and (b)] 
and with that due to finite coupling energy $\varepsilon_{\M}$
 [lower panels (c) and (d)], 
we find that the non-local signal in ${\rm Re}[{S_{\tL\tR}(\omega)}]$ 
arising from $U$ is significantly stronger--about $50$ to$100$ times
larger for the parameters considered.
%
 %
In practice, $\varepsilon_{\M}$ is determined by the
length of the Majorana wire.
For a topological nanowire with a length of $1–2 \mu m$ \cite{Sar12220506},
$\varepsilon_{\M}$
 is at most tens of
${\rm \mu eV}$ much smaller than the typical Coulomb interaction $U$ 
in a QD (on the order of ${\rm meV}$ \cite{Ric20165104}).
In the topologically relevant long-wire limit, $\varepsilon_{\M}\rightarrow0$ in general.
Therefore, Coulomb interaction between QDs and Majorana wire
provides an efficient mechanism for generating nonzero cross correlation.

 

\section{Summary}
\label{thsum}

 We have investigated quantum transport in a hybrid QD-MZM system, 
 taking zero Majorana coupling energy 
($\varepsilon_{\M}=0$) to correspond to the topologically relevant 
  long-wire limit.  
Using the DEOM approach, we accurately evaluated
  both the transient current and the current-current cross-correlation noise spectrum.
Our results
demonstrated that the Coulomb interaction between QDs and a Majorana wire provides 
an efficient mechanism for generating and probing nonlocal electron teleportation mediated by MZMs.

 In the absence of interactions, 
 destructive interference between the normal and anomalous tunneling channels 
 suppresses teleportation.
 As a result, the transport current through the right 
 and left electrodes become independent of each other,
 yielding a vanishing cross-correlation noise.
  A finite Coulomb interaction lifts this channel degeneracy, 
  establishing strong nonlocal correlations between the dots. 
  Consequently, the transient current of the right electrode 
 becomes sensitive
  to the coupling strength between the MZM and left dot.
  This behavior manifests as a robust, experimentally accessible signal 
  in the cross-correlation noise spectrum--- 
  significantly stronger than the signal induced 
  by a finite Majorana coupling energy. 
  Our work highlights Coulomb interaction 
  as a potent and practical control parameter 
  for realizing and detecting Majorana-mediated nonlocal transport in the topologically ideal long-wire limit.

\acknowledgments
We acknowledge helpful discussions with
 Prof. Xin-Qi Li. 
  The support from the 
  National Key Research and Development Program of China (Grant No.  2024YFA1408900)
   is acknowledged.



\end{document}